\documentclass[twocolumn,aps,prl,superscriptaddress,footinbib,amsmath,amssymb,aps,floatfix]{revtex4-1}

\usepackage{bm}
\usepackage{graphicx}
\usepackage{braket}

\begin{document}
\title{Imaging and addressing of individual fermionic atoms in an optical lattice}
\author{G.\ J.\ A.\ Edge}
\author{R.\ Anderson}
\author{D.\ Jervis}
\altaffiliation[Current address: ]{Aerodyne Research Inc., Billerica, MA}
\author{D.\ C.\ McKay}
\altaffiliation[Current address: ]{IBM T.\ J.\ Watson Research Center, Yorktown Heights, NY}
\author{R.\ Day}
\author{S.\ Trotzky}
\affiliation{Department of Physics, University of Toronto, M5S 1A7 Canada}
\author{J.\ H.\ Thywissen}
\affiliation{Department of Physics, University of Toronto, M5S 1A7 Canada}
\affiliation{Canadian Institute for Advanced Research, Toronto, M5G 1Z8 Canada}

% ------------------------------- Abstract ------------------------------

\date{{\today}}
\begin{abstract}
We demonstrate fluorescence microscopy of individual fermionic potassium atoms in a 527-nm-period optical lattice. Using electromagnetically induced transparency (EIT) cooling on the 770.1-nm D$_1$ transition of $^{40}$K, we find that atoms remain at individual sites of a 0.3-mK-deep lattice, with a $1/e$ pinning lifetime of $67(9)\,\rm{s}$, while scattering $\sim 10^3$ photons per second. The plane to be imaged is isolated using microwave spectroscopy in a magnetic field gradient, and can be chosen at any depth within the three-dimensional lattice. With a similar protocol, we also demonstrate patterned selection within a single lattice plane. High resolution images are acquired using a microscope objective with 0.8 numerical aperture, from which we determine the occupation of lattice sites in the imaging plane with 94(2)\% fidelity per atom. Imaging with single-atom sensitivity and addressing with single-site accuracy are key steps towards the search for unconventional superfluidity of fermions in optical lattices, the initialization and characterization of transport and non-equilibrium dynamics, and the observation of magnetic domains.
\end{abstract}
\maketitle

% ------------------------------- Intro ------------------------------

Ultracold fermionic atoms in an optical lattice realize an impurity-free analog of electrons in crystalline materials, with full control of parameters such as interaction strength, dimensionality, and tunneling \cite{Jaksch:2005go,Lewenstein:2007hr}. Furthermore, ultracold systems can study many-body physics in scenarios currently inaccessible to materials, such as gauge fields equivalent to thousands of Tesla \cite{Gerbier:2010ho,Aidelsburger:2013,Miyake2013}, interactions at the unitary limit \cite{Zwerger:2012}, and quantum many-body physics far from equilibrium \cite{Lamacraft:2011wd}. With sufficient control and probes, these experiments can be considered analog quantum simulations \cite{Bloch:2012ty,Georgescu:2014bg}. However, two important tools have been lacking: imaging and addressing fermionic atoms at the single-site and single-atom level \cite{Georgescu:2014bg}. When applied to bosonic atoms, these tools have already been dramatically successful \cite{Nelson:2007ks,Gemelke:2009ja,Bakr:2009bx,Sherson:2010hg,Bakr:2010gd,Weitenberg:2011vm,Fukuhara:2013ki,Ott2015,Miranda:2015wi,Islam:2015ve,Yamamoto:2015td}.

High-resolution imaging and manipulation of ultracold fermions solves several outstanding problems at once. First, in-situ spatial probes directly reveal the order parameter of insulating phases, magnetic domain formation, and other correlations inaccessible in time-of-flight imaging \cite{Sherson:2010hg,Bakr:2010gd,Islam:2015ve}. Second, an ensemble of density distributions provides a direct measure of entropy \cite{Sherson:2010hg,Bakr:2010gd}, extending thermometry of lattice fermions \cite{McKay:2011fc}. Third, manipulation of atoms with single-site precision can initiate dynamics \cite{Weitenberg:2011vm,Fukuhara:2013ki}, project or remove disorder \cite{Bakr:2010gd}, and selectively remove high entropy atoms to perform in-situ cooling \cite{Ho28042009,Bernier:2009jz}.

\begin{figure}[b!]
\includegraphics[width=0.45\textwidth]{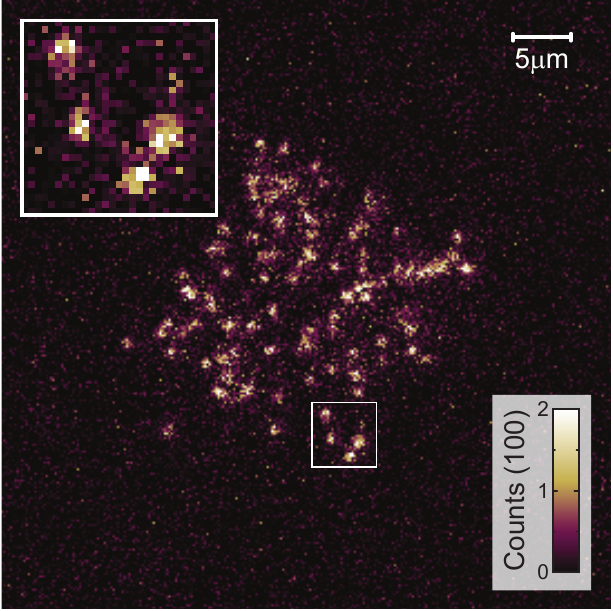}
\caption{High-resolution fluorescence image of fermionic potassium atoms in a single sparsely filled plane of a 527-nm-period cubic optical lattice. The false-color scale indicates the number of counts recorded by and electron-multiplying CCD camera, where one photoelectron corresponds to 16 counts. Atoms outside of a $40\times40$ site box have been removed using the addressing protocol described in the main text. In the inset, one can clearly discern individual atoms. In this 2.6-s-long exposure, an average of $\sim$160 photons are detected per atom.}
\label{fig:image}
\end{figure}

This year, five research groups have succeeded in imaging single fermions in an optical lattice: three using Raman sideband cooling \cite{Cheuk:2015jr,Parsons:2015ck,BlochPC} and two using EIT cooling \cite{Haller:2015hi}, including the results reported in this Article. Our approach is distinguished by a unique imaging configuration, and takes a further step by implementing three-dimensional spatial addressing, which is used here for selective removal of atoms from the lattice. Figure~\ref{fig:image} illustrates these abilities with a high-resolution image of $^{40}$K atoms sparsely filling a selected 40-site $\times$ 40-site $\times$ 1-site volume.

%---------------------------------------------------------------------------------------------------------------------
%----------------------------------------------------------------------------------------------------------------------------

%---------------------------------------------------------------------------------------------------------------------
\begin{figure}[t!]
\includegraphics[width=0.5\textwidth]{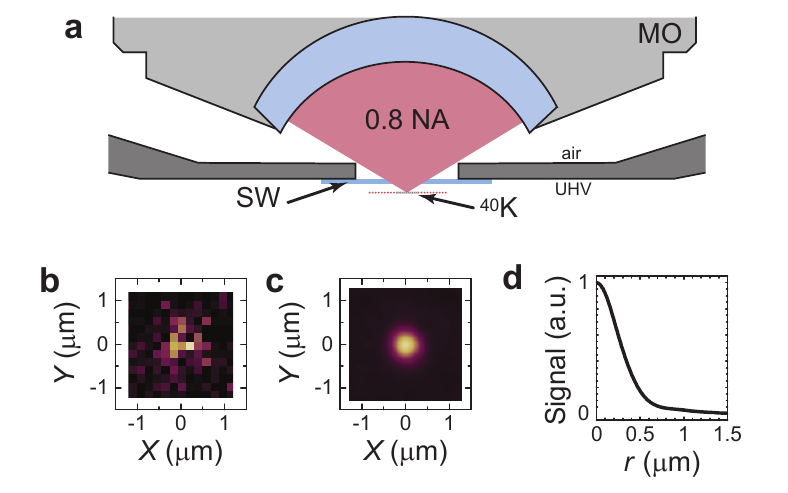}
\caption{Microscopy. (a) A sapphire window (SW) with a thickness of 200$\,\mu$m and a clear aperture of 5\,mm separates UHV from air. This allows for the placement of a microscope objective (MO) with NA $= 0.8$ and 3\,mm working distance outside the vacuum with its focal plane on the inside. (b) Single-shot image of a single atom. With image magnification of $78 \times$, each camera pixel has a virtual size of $192$\,nm. (c) Averaging the signal from 200 isolated atoms in the imaging plane, yields the effective PSF of the imaging system.  The coordinates $X$ and $Y$ are aligned with the principal axes of the images. (d) The radial average of the PSF has a FWHM of 0.60(1)\,$\mu$m. Note that this effective PSF includes not only optical effects but also the finite size of lattice orbitals and drifts of the lattice during exposure. The color scale in (b) and (c) is the same as in Figure~\ref{fig:image}.}
\label{fig:setup}
\end{figure}
%----------------------------------------------------------------------------------------------------------------------------

At the heart of our apparatus is a microscope objective with a numerical aperture (NA) of 0.8, placed outside of an ultra-high vacuum (UHV) chamber, 2.0\,mm above a 200-$\mu$m-thick sapphire window [see Fig.~\ref{fig:setup}(a)]. The focal plane of the imaging system is located inside the vacuum, 0.8\,mm beyond the thin window. Sapphire is sufficiently hard that this thin substrate can sustain atmospheric pressure with a clear aperture of 5\,mm. At the same time, it contributes less spherical aberration than a standard millimeter-thick viewport, since spherical aberration scales as the cube of thickness. The effective point spread function (PSF) of the full imaging system is shown in Fig.~\ref{fig:setup}(c,d) as the average over images of 200 isolated single atoms, centered to sub-pixel precision. Its full width at half maximum (FWHM) of 0.60(1)\,$\mu$m is larger than the diffraction limit of 0.5\,$\mu$m, yet small enough to reconstruct the lattice occupation.

Production of ultracold samples begins with the laser cooling and trapping of fermionic $^{40}$K and bosonic $^{87}$Rb atoms in a glass cell, followed by their transport to a titanium chamber through a succession of magnetic traps. After evaporative and sympathetic cooling to 7\,$\mu$K in a large-volume plugged quadrupole trap, both species are loaded into a crossed optical dipole trap at the focal point of the microscope objective. Further evaporation in the optical dipole trap typically results in degenerate Fermi gases of $2\times10^4$ $^{40}$K atoms at $\leq 0.2\,\mathrm{\mu K}$. After removal of the $^{87}$Rb atoms, this sample is loaded non-adiabatically into a three-dimensional (3D) simple cubic optical lattice, giving a sparse occupation of lattice sites. Each crystal axis is formed by a laser beam at $\lambda_L=1053.6$\,nm: two horizontal beams are retro-reflected in a standard way using mirrors outside of the vacuum system, and the vertical beam is retro-reflected from the sapphire window, which is coated to provide high reflectance at $\lambda_L$. To prevent interference between axes, the beams are cross-polarized and have a relative frequency detuning. The resulting potential is a separable sinusoid of period $\lambda_L/2$. For depths on the order of 10 $E_{RL}$ (where $E_{RL} = h^2 / 2 M \lambda_L^2 \approx k_B \times 216$\,nK and $M$ is the atomic mass and $k_B$ is the Boltzmann constant), such a lattice can be used to explore the physics of fermions in the Hubbard regime \cite{Bloch:2012ty,Georgescu:2014bg}. A much deeper ($\geq 800\,E_{RL}$) lattice 
is used during site-resolved imaging to pin each atom to a single site, with typical harmonic trapping frequencies along the $(x,y,z)$ lattice axes of $2 \pi \times (250,300,250)$\,kHz.

%---------------------------------------------------------------------------------------------------------------------
\begin{figure}[b!]
\includegraphics[width=0.5\textwidth]{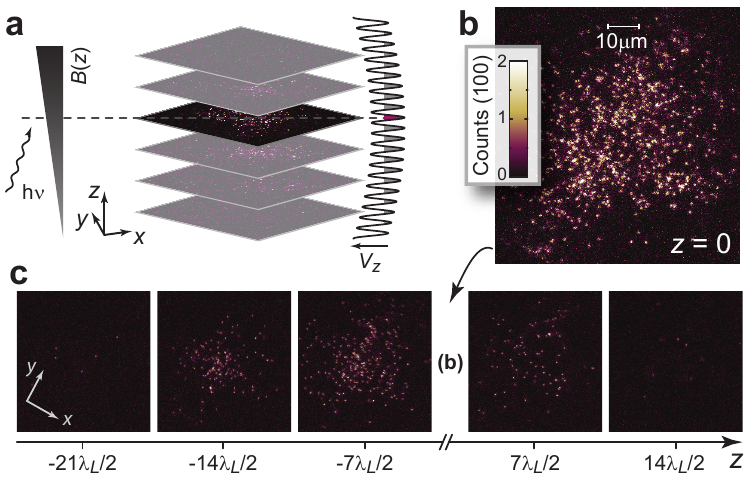}
\caption{Selection of lattice sites. 
(a) Plane selection occurs in three steps. Atoms to be preserved are first transferred from $\ket{F,m_F}=\ket{9/2,-9/2}\equiv\ket{g1}$ to $\ket{7/2,-7/2}\equiv\ket{g2}$ using microwaves at frequency $2 \pi \, \nu$, in the presence to a magnetic gradient $B(z)$. Here $F$ is the total angular momentum and $m_F$ is the magnetic quantum number.  At $105\,{\rm G/cm}$, the 5.5\,mG field difference across $\Delta z =\lambda_L/2$ translates to $\Delta \nu = 14$\,kHz between adjacent planes.
Atoms remaining in $\ket{g1}$ are removed from the lattice with a resonant laser beam operating on the D$_2$ transition, before the atoms shelved in $\ket{g2}$ are brought back to $\ket{g1}$. This selects a single $xy$ plane of the lattice, as shown in (b). Patterning in the $x$ or $y$ plane (see Fig.~\ref{fig:image}) follows a similar procedure. (c) Varying the microwave frequency range chosen to isolate a plane allows for selection of any plane within the bulk 3D lattice.}
\label{fig:select}
\end{figure}
%----------------------------------------------------------------------------------------------------------------------------

In order to provide well resolved images of atoms in the optical lattice, unobscured by the fluorescence of atoms located outside of the focal plane of the objective, a single $xy$ plane of the lattice is isolated prior to imaging. With atoms pinned in the lattice, the desired sites are selected by internal state manipulation in the presence of a magnetic field gradient [see Fig.~\ref{fig:select}(a)] \cite{Karski:2010gk}. The gradient is created by quadrupole coils whose axis of symmetry is along $z$. To select atoms in the $xy$ focal plane of the microscope objective, we apply a $40$\,G field along $z$, and an amplitude-shaped microwave pulse is swept $\pm 5\,{\rm kHz}$ around the resonance frequency of the target lattice plane. This changes the internal state of the atoms with 95\% fidelity, currently limited by our Rabi frequency. Spurious spin-flips in the neighboring planes, $\pm\lambda_L/2$ away, is fully suppressed. After state manipulation, unwanted atoms are removed from the lattice with a 10-ms-long resonant light pulse which has no effect on the transferred atoms. Finally, the selected atoms are returned to the original internal state. After performing this sequence of microwave and optical pulses a second time, $>99$\% of the unwanted atoms have been removed while $90$\% of the desired atoms are preserved.

Patterning along $x$ and $y$ is also possible with this protocol, if the applied bias field is oriented along $x$ or $y$. For example, Fig.~\ref{fig:image} shows the a square pattern selected from the middle of a single lattice plane. This pattern is used for all analyses described below to reduce the effects of lattice inhomogeneity. In-plane patterning has also been demonstrated using projected optical potentials \cite{Bakr:2010gd,Zimmermann:2011it,Weitenberg:2011vm}.

Site-resolved reconstruction of single atoms requires that each atom remains pinned in the lattice while scattering photons. Although direct absorption imaging with short light pulses has been demonstrated for Yb \cite{Miranda:2015wi}, it is not viable for sub-micron single-atom microscopy of alkali atoms. Instead, light scattering must be accompanied with laser cooling. Since red-detuned D$_2$ molasses \cite{Nelson:2007ks,Bakr:2009bx,Sherson:2010hg} is compromised in $^{40}$K due to the inverted hyperfine structure of the  $4P_{3/2}$ excited state, we explored in-situ cooling on the $4S_{1/2}  \to 4P_{1/2}$ (D$_1$) transition at 770.1\,nm in $^{40}$K. Unlike for D$_1$ cooling in free space \cite{Fernandes:2012cr,Salomon:2013ka,Nath:2013fc} or in weak traps \cite{Salomon:2013eu,Burchianti2014} where a Sisyphus mechanism creates a grey molasses \cite{Grynberg:1994ue,*Weidemuller:1994vz,*Boiron:1995tt}, we observe that a polarization gradient is not essential for cooling in a deep lattice. Instead, dark-state coherence establishes an EIT window that suppresses carrier scattering, while creating an absorption resonance at the red trap sideband, thereby cooling the tightly bound atoms \cite{Morigi:2000in,*Morigi:2003cv,*Roos:2000ch,*Mucke:2010gh,Haller:2015hi}. Multicolor Raman sideband cooling realizes a similar mechanism \cite{Vuletic:1998ug,Han:2000vg,Patil:2014jv}, and has also been used for the site-resolved imaging of fermionic atoms \cite{Parsons:2015ck,Cheuk:2015jr}.

%---------------------------------------------------------------------------------------------------------------------
\begin{figure}[b!]
\includegraphics[width=0.5\textwidth]{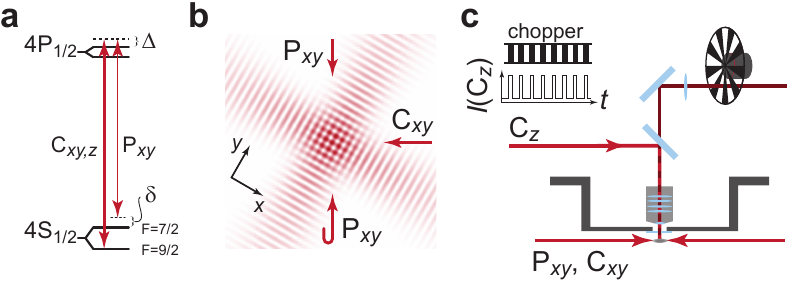}
\caption{Cooling and background reduction. (a) Coupling (C) and probe (P) beams are blue-detuned ($\Delta > 0$) from the $D_1$ transition. (b) Laser cooling in the $xy$-plane involves a retro-reflected probe beam (${\rm P}_{xy}$) and a coupling beam (${\rm C}_{xy}$), both with an angle of $30^\circ$ relative to the nearest lattice beam. (c) An additional coupling beam ${\rm C}_{z}$ passes downward through the sapphire window and extends the cooling to the $z$ direction. Background light from back-scattering of the ${\rm C}_{z}$ beam is minimized by pulsing it synchronously with a chopper wheel blocking all light falling on the EMCCD camera.}
\label{fig:cooling} 
\end{figure}
%----------------------------------------------------------------------------------------------------------------------------

Figure~\ref{fig:cooling} describes our implementation of EIT cooling. Two ``coupling'' beams (C$_{xy}$ and C$_z$) are near the  $4S_{1/2} (F=9/2) \to 4P_{1/2} (F'=7/2)$ transition, while a weaker ``probe'' beam (P$_{xy}$) is near the  $4S_{1/2} (F=7/2) \to 4P_{1/2} (F'=7/2)$ transition with differential detuning of $\delta$ from the Raman resonance across ground states. The common-mode detuning $\Delta$ from the $F'=7/2$ state of all beams depends on the depth of the optical lattice due to the Stark shift, with larger lattice depth corresponding to smaller $\Delta$. For atoms in the center of the lattice at the depth used for imaging, the Stark shift of the D$_1$ transition is measured to be $2 \pi \times 68$\,MHz, and our cooling beams are detuned by $\Delta = 2 \pi \times 36\,$MHz. The beam geometry is shown in Fig.~\ref{fig:setup}(b). The C$_z$ beam has $2\,\mu$W, P$_{xy}$ has $2\,\mu$W and is retroreflected, and C$_{xy}$ has $40\,\mu$W, providing Rabi frequencies of $2 \pi \times 2.3\,$MHz, $2 \pi \times 1.3\,$MHz, and $2 \pi \times 4.2\,$MHz respectively. Applying these beams scatters photons from the trapped atoms, while the EIT cooling mechanism prevents the atoms from heating out of the lattice sites. Time-of-flight expansion from a 3D lattice after band mapping shows that the majority of atoms remain in the ground vibrational band during imaging.

In order to detect the light scattered from single atoms, it is crucial to eliminate background light. Stray light from the lattice beams can be filtered spectrally, whereas the D$_1$ light cannot. We reduce background light from the horizontal beams C$_{xy}$ and P$_{xy}$ with careful beam shaping and alignment, but background scattered light from the vertical cooling beam C$_z$ is unavoidable. The $2\,\mu$W of C$_z$ is roughly $10^{11}$ times more powerful than the $\sim$20\,aW signal of a single atom, and we find that attenuating the C$_z$ background scattering with polarization optics and spatial filters is insufficient in our setup.

Instead, we find that a pulsed cooling method can reduce background light to the level of a single photon per pixel during a 2.6\,{\rm s} exposure. Our approach is depicted in Fig.~\ref{fig:cooling}(c): a chopping wheel blocks all light incident on the camera for 50\% of the exposure, and the beam C$_z$ is  applied for $35$\% of the exposure, during periods for which the camera is fully shielded. When unblocked, the camera collects light scattered by atoms from the P$_{xy}$ and C$_{xy}$ beams, which also leads to heating of the uncooled vertical degree of freedom. For sufficiently low scattering rates (discussed below), we observe that long pinning times are still attainable with this pulsed cooling method. The time-averaged fluorescence signal is however halved, unlike in schemes where vertical cooling entails a distinguishable wavelength \cite{Haller:2015hi,Cheuk:2015jr}. Including chopping, a 20\% collected solid angle, quantum efficiency of the camera, and additional transmission losses, we estimate a net detection efficiency of 7\% for scattered photons. Together with our measured photoelectron signal per atom, this collection efficiency implies that the scattering rate is $\approx 900\,$s$^{-1}$ during EIT cooling.

\begin{figure}[tb!]
\includegraphics[width=0.5\textwidth]{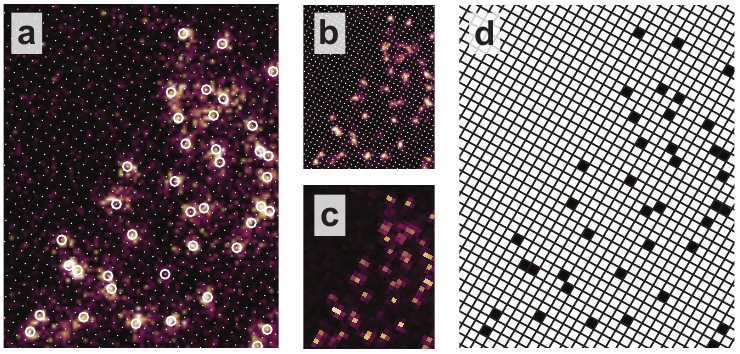}
\caption{Reconstruction of the lattice occupation. (a) Magnified subimage of Fig.~\ref{fig:image}. (b) Deconvolution of the image with the PSF [see Fig.~\ref{fig:setup}(c)] gives a sharpened image which we use to pin the lattice grid. (c) A lattice-site binned image is then used to select potentially occupied sites via a threshold. (d) The final best-fit digitization is determined by comparing all possible occupation patterns for the selected sites and their immediate neighbors with the sharpened image. Those sites that were found to be occupied are also marked with white circles in (a).}
\label{fig:digit}
\end{figure}

Images such as Fig.~\ref{fig:image} reveal the binary filling of all lattice sites in the selected region, with the help of additional information about the lattice periodicity and the PSF. From a number of similar images, we determine the orientation and apparent period of the optical lattice through evaluation of the relative positions of more than 2000 isolated atoms. Comparing to the known lattice spacing of $\lambda_{L}/2$ yields the magnification of our imaging system ($78\times$). With the lattice angles and magnification determined, we can reconstruct the lattice occupation $\in\{0,1\}$ from each fluorescence image. We expect the apparent lattice occupation to be parity sensitive due to light-assisted collisions \cite{Bakr:2009bx,Sherson:2010hg}, however, our average filling is $\ll 1$ atom per site, so that occupancies larger than one are rare. Figure~\ref{fig:digit}(a-d) illustrates the steps taken by our reconstruction algorithm to digitize a raw fluorescence image via sharpening and site-binning.

\begin{figure}[t]
\includegraphics[width=0.5\textwidth]{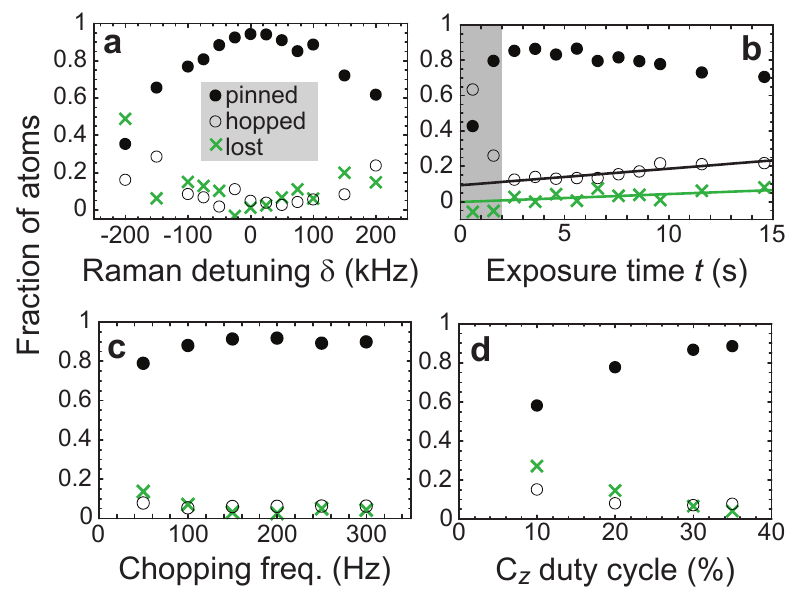}
\caption{Fidelity. Plots show the fraction of atoms in the first digitized image that are pinned (filled circles), hop to a different site (open circles), or are lost completely (crosses) in the successive image. Unless otherwise indicated, $\delta=0$, the C$_z$ chopping rate is 100\,Hz, the C$_z$ duty cycle is 35\%, and the exposure time is 2.6\,s with 0.4\,s between exposures. Maximal pinning occurs for (a) Raman detuning $\delta \approx 0$; (b) exposure time between 2\,s and 5\,s; (c) a chopping frequency that is $> 100$\,Hz; and (d) a duty cycle of C$_z$ that is at least 30\%. Lines shown in (b) are $0.002(28) + 0.004(3)\,t$ for loss fraction  and $0.07(2) + 0.011(2)\,t$ for hopping fraction, and fit only to points at $t>4$\,s.}
\label{fig:fidel}
\end{figure}

The fidelity of imaging and reconstruction is assessed by comparing digitized images from two sequential exposures of the same arrangement of atoms. Exposures are separated by the 0.4\,s required for camera read-out, during which atoms are still laser-cooled. By counting the number of atoms in the second digitization that either appear at an empty site or disappear from an occupied one, we calculate the fraction of atoms which are pinned, hop to a different site, or are lost completely in the second exposure. Figure~\ref{fig:fidel} shows these measures versus several critical imaging parameters.

%---------------------------------------------------------------------------------------------------------------------
%----------------------------------------------------------------------------------------------------------------------------

%---------------------------------------------------------------------------------------------------------------------
%----------------------------------------------------------------------------------------------------------------------------

Figure~\ref{fig:fidel}(a) shows that optimal cooling is observed for $\delta = 0\,$kHz, as was found in prior work \cite{Morigi:2000in,*Morigi:2003cv,*Roos:2000ch,*Mucke:2010gh,Haller:2015hi}. Here, the dressed ground state is maximally dark to elastic scattering, and inelastic scattering is biased towards red (cooling) transitions. Figure~\ref{fig:fidel}(b) shows that long exposures are possible with high fidelity. At short ($<2$\,s) exposure times, the apparent hopping fraction is high, due to errors in reconstruction with insufficient signal. However at longer exposure times, reconstruction errors are negligible, and loss and hopping approach constant rates of 0.4(3)\%s$^{-1}$ and 1.1(2)\%s$^{-1}$, respectively. This loss rate is consistent with a $1/e$ trap lifetime of $>$200\,s, and a pinned fraction lifetime of 67(9)\,s.

Figures~\ref{fig:fidel}(c,d) evaluate the conditions under which a modulated C$_z$ provides sufficient cooling to the vertical degree of freedom to maintain fidelity. Reducing the chopping frequency below 100\,Hz results in a decrease in the fraction of atoms that are pinned to their sites. Fig~\ref{fig:fidel}(d) shows that the rate of loss and hopping increases if the duty cycle of C$_z$ is lower than 20\%, at a chopping rate of 100\,Hz. Thus, a high pinned atom fraction is observed with $\gtrsim$3\,ms cooling pulses. Combined with the inferred scattering rate, this suggests that approximately six photons can be scattered between vertical cooling cycles. 

In optimal conditions, we find that 94(2)\% of atoms stay pinned to the same lattice site in a sequence of two images. This is comparable to performance reported in Refs.~\onlinecite{Parsons:2015ck,Haller:2015hi,Cheuk:2015jr}, where pinning fidelity between successive images ranged from 92\% to 95\%. The fraction of atoms lost in the second image can be as low as 2(1)\%. The rest of the 6(2)\% of atoms that do not stay pinned either hop or are incorrectly reconstructed in the first or second frame. The optimal exposure time must compromise between signal and hopping during imaging, as shown in Fig.~\ref{fig:fidel}(b). This optimum will also depend on lattice filling fraction, since true hopping will become more problematic at high density, where such events will more likely eject a pair of atoms due to light-assisted collisions \cite{Nelson:2007ks,Bakr:2009bx,Sherson:2010hg}.

In sum, we demonstrate single-atom imaging and selection of fermionic potassium atoms in a far-detuned optical lattice. A crucial realization is that pulsed EIT cooling on the D$_1$ transition of $^{40}$K can provide sufficient signal and lifetime for fluorescence microscopy. By reducing background light below the level of a single photon per pixel, the occupation of each lattice site in the imaging plane can be determined with several hundred photons collected per pinned atom.
We furthermore demonstrate spectroscopic selection of sites, which has two implications. First, spatially selective removal of atoms is essential to implement the cooling mechanism proposed in Refs.~\onlinecite{Ho28042009,Bernier:2009jz}. Second, although the microscope can only characterize a single 2D plane, it can be any plane of the 3D optical lattice [see Fig.~\ref{fig:select}(c)]. This enables the tomographic exploration of 3D physics, which is crucial to model materials such as cuprates, in which ``c-axis'' tunneling is responsible for long-range order in anti-ferromagnetic and superconducting phases 
\cite{Chakravarty:1988jg,*Chakravarty:1989he,Kastner:1998el,Li:2007ka,*Berg:2007kn}. Even for materials, scanning probes can only measure the surface, while here the imaging plane can be submerged in the bulk.
Our results, along with those of Refs.~\onlinecite{Parsons:2015ck,BlochPC,Haller:2015hi,Cheuk:2015jr}, are crucial steps towards the study of strongly correlated phases of fermionic atoms in optical lattices.

\acknowledgments{We thank S.\ Heun, C.\ Kierans, T.\ Maier, J.\ McKeever, J.\ Metzkes, D.\ Nino, D.\ Fine, K.\ Pilch, M.\ Sprague, J.\ Sutton, Chen Ge Qu, C.\ Veit, M.\ Yee, and Tout Wang for experimental assistance, and M.\ Greiner, C.\ Luciuk, D.\ Weiss, C.\ Weitenberg, and M.\ Zwierlein for helpful discussions. This work was supported by AFOSR under FA9550-13-1-0063, by CFI, by the DARPA OLE program, and by NSERC.}

\bibliography{bibMicroscopeV10}

\end{document}